\documentclass[fleqn]{article}
\usepackage[english]{babel}
\usepackage{a4wide}
\usepackage{latexsym}
\usepackage{theorem}
\usepackage{url}
\usepackage[final]{graphics}
\usepackage{amsmath,amssymb}
\usepackage{amsfonts}
\usepackage{array}
\usepackage{calc}
\usepackage{xspace}
\usepackage{color}
\usepackage{epsfig}
\usepackage{subfigure}
\usepackage{float}
\usepackage{hyperref}
\usepackage{stmaryrd}
\usepackage{color}
\usepackage{mathtools}
\usepackage{tikz}
\usetikzlibrary{calc,arrows}
\usetikzlibrary{decorations.pathreplacing}

\newcolumntype{L}{>{$}l<{$}}
\newcolumntype{C}{>{$}c<{$}}
\newcolumntype{R}{>{$}r<{$}}

{\begin{trivlist}
\item\begin{tabular}{@{}>{\bf}p{2.3em}L@{\ }L@{\ }L@{\ }L@{\ }L@{\ }L@{\ }L@{\ }L}}%
{\end{tabular}\end{trivlist}}

\newcommand{\Real}{\mathbb{R}}
\newcommand{\Nat}{\mathbb{N}}
\newcommand{\Int}{\mathbb{Z}}
\newcommand{\Pos}{\mathbb{N}^+}
\newcommand{\Bool}{\mathbb{B}}
\newcommand{\dataeq}{\approx}
\newcommand{\true}{\textit{true}}
\newcommand{\false}{\textit{false}}
\usepackage{todonotes}

\usepackage{wrapfig}

\newcommand{\textitprob}{p}  

\begin{document}
\title{On Woolhouse's Cotton-Spinning Problem}
\author{{Jan Friso~Groote and Tim A.C.~Willemse}\\
\small Department of Mathematics and Computer Science,
\small Eindhoven University of Technology\\
\small P.O.~Box 513, 5600 MB Eindhoven, The Netherlands\\
\texttt{\small \{J.F.Groote, T.A.C.Willemse\}@tue.nl}}
\date{}
\maketitle
\begin{abstract}
\noindent%
In 1864 W.S.B.~Woolhouse formulated the Cotton-Spinning problem \cite{Woolhouse64}. 
This problem boils down to the following. A piecer works at a spinning mule
and walks back and forth to repair broken threads. The question is how far the piecer 
is expected to walk when the threads break at random. 
This problem can neatly 
be solved using process modelling and quantitative model checking, showing that 
Woolhouse's model led to an overestimation of the walking distance.  
\end{abstract}
\section{Introduction}

%

\begin{wrapfigure}{r}{0.4\textwidth}
  \vspace{-20pt}
  \begin{center}
  \includegraphics[width=0.38\textwidth]{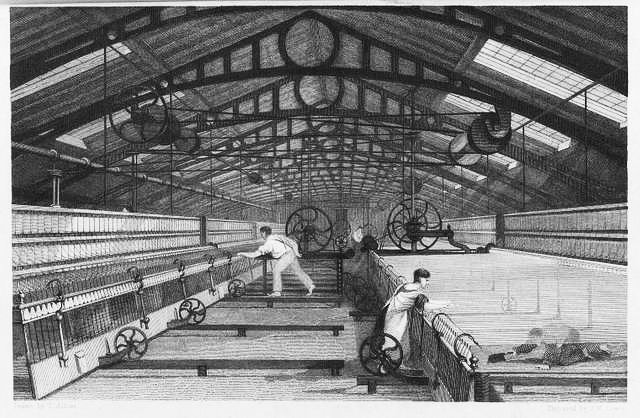}
  \end{center}
  \vspace{-20pt}
  \caption{Spinning mules in a factory}
  \label{fig:baines}
  \vspace{-10pt}
\end{wrapfigure}

The Industrial Revolution, which started in Great Britain in the 18\textsuperscript{th} century, saw the rise of mechanised factory systems and efficient manufacturing processes.
Textile production was among the first to take an industrial spin, with machines called \emph{spinning mules} for spinning cotton and other fibres being used extensively.
These mules, which could be operated by a \emph{minder} -- also known as a mule spinner -- 
and two \emph{piecers}, allowed for a huge reduction in required labour and led to significant cost reductions of spinning.
The machines consisted of a carriage that was able to carry up to 1,320 spindles and could be up-to 46 meters long. 
A carriage would move back and forth over a distance of 1.5 meters, four times a minute \cite{Catling86}.
The mule would carry spindles that would both twist and take up the spun thread as the carriage would be moving. 
See Figure \ref{fig:baines} for an impression of a factory with spinning mules (the picture originates from~\cite{baines1835}).

Piecers, typically young girls or boys, would repair the sporadic yarn breakages.
They would walk with and along the mule as it moved, catch ending or broken threads and piece them.
The quality of the roving greatly impacted the number of yarn breakages, with 5 to 6 breakages a minute being typical of machines in the early 20\textsuperscript{th} century.
While repair could be done in a few seconds, requiring only a slight rolling of the forefinger against the thumb, it had to be done while the mule was moving. 

The actuary W.S.B.~Woolhouse, seemingly concerned with the welfare of the factory workers, set out to investigate the average distance piecers walked in a given day.
After taking note of the various dimensions of the mules typically found in the greater Manchester district, the number of strokes of a machine per minute, and the expected number of breakages of the threads at each stroke, he computed that distances travelled in excess of 30~miles ($>$45km) would not be uncommon.

In 1864 Woolhouse presented the problem from a mathematical point of view as follows in a publication in
\emph{The Assurance Magazine and Journal of the Institute of Actuaries}~\cite{Woolhouse64}.
We henceforth refer to the problem to determine the walking distance of a piecer as \textit{Woolhouse's problem}.
Woolhouse solves the problem using a model where the piecer starts at a uniformly chosen point behind the mule
and calculates the required walking distance during one stroke of the mule with $N$ uniformly distributed
broken threads. In his words  \cite{Woolhouse64}:
\begin{quote}\em
``Supposing $n$ points to be taken promiscuously on a line of a given length, and that a person stationed somewhere on the line is required to proceed to all the points by the shortest route, determine the average distance that he may be expected to travel''.
\end{quote}


As Woolhouse's model only considers one stroke of the mule, the model only describes the actual behaviour of the piecer 
in a limited way, which reduces the adequacy of his results. 
It would be nicer to model the situation where the piecer walks back and forth the mule a number of times. However,
this is tricky to model with classical probability theory. 

The actual behaviour of the piecer can, instead, be very neatly modelled using probabilistic 
process formalisms. In this particular case, we use mCRL2 \cite{GM14}.
Using a recently developed quantitative modal logic~\cite{GW23} the average walking distances can be easily
expressed. By solving the modal formulas on the behavioural models the required results are obtained.

In this paper, in Section \ref{section2}, we first reformulate Woolhouse's model, albeit with a small change, which we believe mends
a minor aberration of the original model. We give an equivalent process description for this model, which additionally enables us to
analyse the situation where a piecer walks back and forth the mule during multiple strokes. 

Woolhouse's approach is not truly natural as he assumes a fixed number $N$ of
broken threads per stroke. Therefore, in Section \ref{section3}, we consider a more natural model where each
thread can break with equal probability. This may lead to the situation that sometimes no threads break, and,
very rarely, every thread breaks. This model is computationally expensive because exponentially many breakage 
patterns must be considered. To cope with the complexity of this new model, 
we introduce an alternative model that is more optimal, yet remains probabilistically bisimilar to the expensive model. Using this equivalent model we can make nice assessments of the average 
walking distances of the piecers. Our analysis, reflected upon in the last section, indicates that
Woolhouse's model tends to overestimate the walking distances.

We provided an analysis in this paper using mCRL2 and quantitative modal formulas. 
However, there are other probabilistic process tools using which similar analyses can be
done. 
Typical tools are Storm \cite{Storm}, Prism \cite{Prism} and Modest \cite{Modest}.

\section{Woolhouse's model}
\label{section2}

Woolhouse provides an answer to his problem in \cite{Woolhouse64}. We explain Woolhouse's approach but give our own 
probabilistic model 
which deviates from Woolhouse's formulation as he slightly miscounts the number of breakage patterns.
But the models are very close, especially for large mules. The probabilistic model is quite technical and
is not essential to understand the rest of the paper. 
Subsequently, we provide a model in mCRL2, analysed by quantitative modal formulas, and
show that this model exactly coincides with our probabilistic model. The mCRL2 model is more suitable
to study the actual distance the piecer walks.

\subsection{A probabilistic model}
\label{sec:probmod}
\begin{figure}[t]
\begin{center}
\begin{tikzpicture}
\draw [-, thick] (0,4.6) --( 10,4.6);
\draw [-, thick] (0,4.6) --( 0,4.4);
\draw [-, thick] (1,4.6) --( 1,4.4);
\draw [-, thick] (2,4.6) --( 2,4.4);
\draw [-, thick] (3,4.6) --( 3,4.4);
\draw [-, thick] (4,4.6) --( 4,4.4);
\draw [-, thick] (5,4.6) --( 5,4.4);
\draw [-, thick] (6,4.6) --( 6,4.4);
\draw [-, thick] (7,4.6) --( 7,4.4);
\draw [-, thick] (8,4.6) --( 8,4.4);
\draw [-, thick] (9,4.6) --( 9,4.4);
\draw [-, thick] (10,4.6) --( 10,4.4);
\draw (0.5,4.4) node {$0$};
\draw (10.5,4.4) node {$\textit{width}$};
\draw (5.5,4.4) node {$\textit{pos}$};
\draw [decoration={brace}, decorate] (2,4.7) -- (5,4.7);
\draw (3.5,4.95) node {$\beta$};
\draw (2.5,4.4) node {\small broken};
\draw [->, dashed, thick] (5.5, 4.2) -- (2.5,4.2);
\draw [-, thick] (0,3.4) --( 10,3.4);
\draw [-, thick] (0,3.4) --( 0,3.2);
\draw [-, thick] (1,3.4) --( 1,3.2);
\draw [-, thick] (2,3.4) --( 2,3.2);
\draw [-, thick] (3,3.4) --( 3,3.2);
\draw [-, thick] (4,3.4) --( 4,3.2);
\draw [-, thick] (5,3.4) --( 5,3.2);
\draw [-, thick] (6,3.4) --( 6,3.2);
\draw [-, thick] (7,3.4) --( 7,3.2);
\draw [-, thick] (8,3.4) --( 8,3.2);
\draw [-, thick] (9,3.4) --( 9,3.2);
\draw [-, thick] (10,3.4) --( 10,3.2);
\draw (0.5,3.2) node {$0$};
\draw (10.5,3.2) node {$\textit{width}$};
\draw (5.5,3.2) node {$\textit{pos}$};
\draw [decoration={brace}, decorate] (6,3.5) -- (9,3.5);
\draw (7.5,3.75) node {$\beta$};
\draw (8.5,3.2) node {\small broken};
\draw [->, dashed, thick] (5.5, 3.0) -- (8.5,3.0);
\draw [-, thick] (0,2.2) --( 10,2.2);
\draw [-, thick] (0,2.2) --( 0,2.0);
\draw [-, thick] (1,2.2) --( 1,2.0);
\draw [-, thick] (2,2.2) --( 2,2.0);
\draw [-, thick] (3,2.2) --( 3,2.0);
\draw [-, thick] (4,2.2) --( 4,2.0);
\draw [-, thick] (5,2.2) --( 5,2.0);
\draw [-, thick] (6,2.2) --( 6,2.0);
\draw [-, thick] (7,2.2) --( 7,2.0);
\draw [-, thick] (8,2.2) --( 8,2.0);
\draw [-, thick] (9,2.2) --( 9,2.0);
\draw [-, thick] (10,2.2) --( 10,2.0);
\draw (0.5,2.0) node {$0$};
\draw (10.5,2.0) node {$\textit{width}$};
\draw (5.5,2.0) node {$\textit{pos}$};
\draw [decoration={brace}, decorate] (3,2.3) -- (5,2.3);
\draw (4,2.55) node {$\beta$};
\draw (3.5,2.0) node {\small broken};
\draw [decoration={brace}, decorate] (6,2.3) -- (9,2.3);
\draw (7.5,2.55) node {$\beta'$};
\draw (8.5,2.0) node {\small broken};
\draw [->, dashed, thick] (5.5, 1.8) -- (3.5,1.8);
\draw [->, dashed, thick] (3.5, 1.7) -- (8.5,1.7);
\draw [-, thick] (0,1.0) --( 10,1.0);
\draw [-, thick] (0,1.0) --( 0,0.8);
\draw [-, thick] (1,1.0) --( 1,0.8);
\draw [-, thick] (2,1.0) --( 2,0.8);
\draw [-, thick] (3,1.0) --( 3,0.8);
\draw [-, thick] (4,1.0) --( 4,0.8);
\draw [-, thick] (5,1.0) --( 5,0.8);
\draw [-, thick] (6,1.0) --( 6,0.8);
\draw [-, thick] (7,1.0) --( 7,0.8);
\draw [-, thick] (8,1.0) --( 8,0.8);
\draw [-, thick] (9,1.0) --( 9,0.8);
\draw [-, thick] (10,1.0) --( 10,0.8);
\draw (0.5,0.8) node {$0$};
\draw (10.5,0.8) node {$\textit{width}$};
\draw (5.5,0.8) node {$\textit{pos}$};
\draw [decoration={brace}, decorate] (2,1.1) -- (5,1.1);
\draw (3.5,1.35) node {$\beta$};
\draw (2.5,0.8) node {\small broken};
\draw [decoration={brace}, decorate] (6,1.1) -- (8,1.1);
\draw (7,1.35) node {$\beta'$};
\draw (7.5,0.8) node {\small broken};
\draw [->, dashed, thick] (5.5, 0.6) -- (7.5,0.6);
\draw [->, dashed, thick] (7.5, 0.5) -- (2.5,0.5);
\end{tikzpicture}
\end{center}
\caption{The four walking patterns of the piecer}
\label{fig:walking_patterns}
\end{figure}
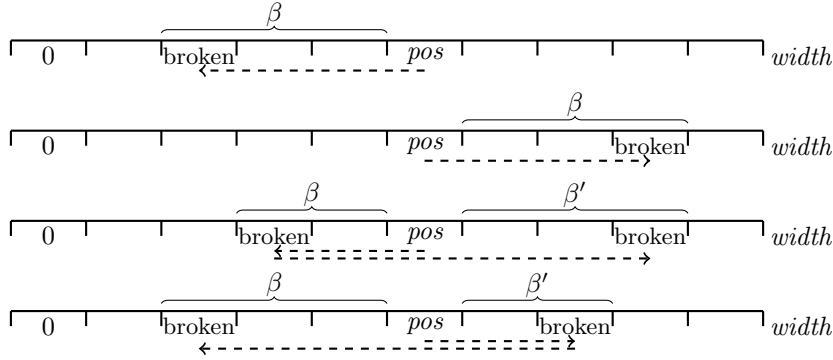

Assume the spinning mule has a width $\textit{width}$ of positions and at exactly $N$ of these positions threads 
break.
The piecer stands at a position $\textit{pos}$ and
walks either to the left, to the right, or in both directions to repair all broken threads. 
Woolhouse uses the letters $a$, $n$ and $P$ for respectively $\textit{width}$, $N$ and $\textit{pos}$. 
The four
situations that can occur are depicted in Figure \ref{fig:walking_patterns}.

We first look at the situation where the piecer only needs to walk a distance $\beta$ to the left
as all broken threads occur at $\textit{pos}$ or at the left of $\textit{pos}$ and the furthest is exactly $\beta$ away. 
This corresponds to the upper diagram in Figure \ref{fig:walking_patterns}.
Note that $\beta\geq N-1$ as all broken threads must be situated at distinct positions.
The probability corresponding to this situation is
\[ \binom{\beta}{N-1}/\binom{\textit{width}}{N}.\]
If we let $\beta$ range from $N-1$ to $\textit{pos}$, we get the expected distance that the piecer needs to walk in this case:
\[\Delta_1=\sum_{\beta=N-1}^\textit{pos} \frac{\beta\binom{\beta}{N-1}}{\binom{\textit{width}}{N}}=
\frac{(\textit{pos}-N+2)(N\,\textit{pos}+N-1)\binom{\textit{pos}+1}{N-1}}{N(N+1)\binom{\textit{width}}{N}}.\]

The probability that all broken threads are at $\textit{pos}$ or at the right of the piecer, and the furthest thread is $\beta$ away, is
the same as above. In this case the expected distance to walk to the right is the following:
\[\Delta_2=\sum_{\beta=N-1}^{\textit{width}-\textit{pos}-1} \frac{\beta\binom{\beta}{N-1}}{\binom{\textit{width}}{N}}=
\frac{(\textit{pos}+\textit{width}-N+1)(a N-N\textit{pos}-1)\binom{\textit{width}-\textit{pos}}{N-1}}{N(N+1)\binom{\textit{width}}{N}}.\]

In situations where the piecer finds a broken thread at the left at distance $\beta$ and at the right at distance $\beta'$ and
all other threads are in between, has probability:
\[ \binom{\beta+\beta'-1}{N-2}/\binom{\textit{width}}{N}.\]
If $\beta\leq \beta'$, this corresponds to the one but lowest diagram in Figure \ref{fig:walking_patterns}. The piecer walks $2\beta+\beta'$. So, the expected walking distance is
\[\Delta_3=\sum_{\beta=1}^\textit{pos}\sum_{\beta'=\max(\beta,N-1-\beta)}^{\textit{width}-\textit{pos}-1}(2\beta+\beta') \binom{\beta+\beta'-1}{N-2}/\binom{\textit{width}}{N}.\]
Reversely, if $\beta> \beta'$, corresponding to the lowest diagram in Figure \ref{fig:walking_patterns}, 
the piecer walks $\beta+2\beta'$. So, the expected walking distance is
\[\Delta_4=\sum_{\beta=1}^{\textit{pos}}\sum_{\beta'=\max(1,N-\beta)}^{\min(\textit{width}-\textit{pos}-1,\beta-1)}
(\beta+2\beta') \binom{\beta+\beta'-1}{N-2}/\binom{\textit{width}}{N}.\]
The total expected walking distance $\Delta=\Delta_1+\Delta_2+\Delta_3+\Delta_4$ as a fraction of the width of the mule is given in Table \ref{table:walking_distance}.
This table is almost the same as the table given in \cite{Woolhouse64} despite the different
calculation. It is calculated with $\textit{width}=10000$.
\begin{table}[h]
\begin{center}
\begin{tabular}{|c|p{1.33cm}|p{1.33cm}|p{1.33cm}|p{1.33cm}|p{1.33cm}|p{1.33cm}|}
\hline
$N$&\multicolumn{6}{@{}c@{}}{
\begin{tabular}{|p{1.33cm}|p{1.33cm}|p{1.33cm}|p{1.33cm}|p{1.33cm}|p{1.33cm}|}
\multicolumn{6}{|p{8cm}|}{\mbox{Relative starting point of the piecer, i.e., $\textit{pos}/\textit{width}$}}\\
\hline
$0.0$&$0.1$&$0.2$&$0.3$&$0.4$&$0.5$\\
\end{tabular}}
\\
\hline
1 &0.5000 & 0.4100 & 0.3400 & 0.2900 & 0.2600 & 0.2500\\
2 &0.6666 & 0.5853 & 0.5360 & 0.5107 & 0.5014 & 0.5000\\
3 &0.7500 & 0.6779 & 0.6524 & 0.6580 & 0.6765 & 0.6876\\
4 &0.8000 & 0.7362 & 0.7302 & 0.7605 & 0.8025 & 0.8251\\
5 &0.8333 & 0.7772 & 0.7871 & 0.8357 & 0.8952 & 0.9272\\
6 &0.8571 & 0.8081 & 0.8313 & 0.8931 & 0.9651 & 1.0046\\
7 &0.8750 & 0.8327 & 0.8670 & 0.9385 & 1.0190 & 1.0646\\
8 &0.8889 & 0.8528 & 0.8966 & 0.9752 & 1.0617 & 1.1121\\
9 &0.9000 & 0.8698 & 0.9216 & 1.0055 & 1.0960 & 1.1505\\
10&0.9091 & 0.8844 & 0.9430 & 1.0309 & 1.1242 & 1.1821\\
\hline
\end{tabular}
\end{center}
\caption{Walking distances relative to the width of the mule (model in Section \ref{sec:probmod}, $\textit{width}=10000$)}
\label{table:walking_distance}
\end{table}

From this table Woolhouse concludes that the piecer must walk substantial distances. In particular he concludes
that the piecer may have to walk more than the length of the spinning mule if the number of broken threads $N$
is high and the piecer starts in the middle. Woolhouse observes that ``when this is the case, the piecer would
do well to walk backwards and forwards alternately the entire distance'' \cite{Woolhouse64}.

This indicates a prime weakness in Woolhouse's analysis. He assumes that the piecer starts either at a fixed position,
or uniformly distributed over all starting points. 
But in reality the piecer will start at the position where he stopped in the previous round which is unlikely
to be uniformly distributed. We provide a model in mCRL2 which allows us to investigate the behaviour of the 
piecer during multiple rounds. 

\subsection{Woolhouse's model in mCRL2}
\begin{figure}[tp]
\begin{center}
\begin{tabular}{|ll|}
\hline
&\\
\textbf{sort} &$\textit{Thread} = \textbf{struct}~ \textit{fine} \mid \textit{broken};$\\
&\\
\textbf{map}&  $N, \textit{initial\_position}, \textit{width}: \Nat;$\\
&     $\textit{count\_broken\_threads},  \textit{lbt}, \textit{rbt}: \textit{List}(\textit{Thread})\rightarrow\Nat; ~~\% \textit{lbt}: \textrm{leftmost broken thread};$\\
&$\hspace{7.07cm}\%\textit{rbt}: \textrm{rightmost broken thread}.$\\
&     $\textit{no\_broken\_thread}: \textit{List}(\textit{Thread})\rightarrow \Bool;$\\
&     $\textit{binomial}: \Nat\times\Nat\rightarrow\Real;$\\
&     $\textit{fac}: \Nat\rightarrow\Nat;$\\
&\\
\textbf{eqn}&$  \textit{width}=20;$\\
&     $N=1;$\\
&     $\textit{initial\_position}=10;$\\
&\\
\textbf{var}&  $n,m:\Nat;$\\
\textbf{eqn}&  $\textit{fac}(0)=1;$\\
&     $n>0 \rightarrow \textit{fac}(n)=\textit{fac}(\max(0,n-1))*n;$\\
&     $\textit{binomial}(n,m)=\textit{fac}(n)/(\textit{fac}(\textit{Int2Nat}(n-m))*\textit{fac}(m));$\\
&\\
\textbf{var}
&     $l: \textit{List}(\textit{Thread});$\\
\textbf{eqn}
     &$\textit{lbt}([])=0;$\\
     &$\textit{lbt}(\textit{fine}{\triangleright}l)=1{+}\textit{lbt}(l);$\\
     &$\textit{lbt}(\textit{broken}{\triangleright} l)=0;$\\
     &$\textit{rbt}(\textit{fine}{\triangleright}l)=1{+}\textit{rbt}(l);$\\
     &$\textit{rbt}\textit{(broken}{\triangleright}l)=\textit{if}(\textit{no\_broken\_thread}(l),0,1{+}\textit{rbt}(l));$\\
     &$\textit{no\_broken\_thread}([])=\true;$\\
     &$\textit{no\_broken\_thread}(\textit{fine}{\triangleright}l)=\textit{no\_broken\_thread}(l);$\\
     &$\textit{no\_broken\_thread}(\textit{broken}{\triangleright}l)=\false;$\\

&     $\textit{count\_broken\_threads}([])=0;$\\
&     $\textit{count\_broken\_threads}(\textit{fine} \triangleright l)=\textit{count\_broken\_threads}(l);$\\
&     $\textit{count\_broken\_threads}(\textit{broken}\triangleright l)=1+\textit{count\_broken\_threads}(l);$\\
&\\
\textbf{act}&$  \textit{threads}: \textit{List}(\textit{Thread});$\\
&     $\textit{walk}:\Int;$\\
&\\
\textbf{proc}&$ \textit{Mule}(\textit{pos}:\Nat)=$\\
&  $     \hspace*{0.5cm}\textbf{dist}~ l{:} \textit{List}(\textit{Thread})[\textit{if}(\#l\dataeq \textit{width} \wedge \textit{count\_broken\_threads}(l){\dataeq} N,$\\
&  $     \hspace*{6cm}
                              1/\textit{binomial}(\textit{width},N),0)].$\\
&   $\hspace{1cm}\textit{threads}(l){\cdot}$\\

&$\hspace{1cm}((\textit{pos}{\leq}\textit{lbt}(l))  {\rightarrow}  \textit{walk}(\textit{rbt}(l)
      {-}\textit{pos}){\cdot}\textit{Mule}(\textit{rbt}(l)) +{}$\\
&$\hspace{1.15cm}(\textit{pos}{\geq}\textit{rbt}(l)) {\rightarrow}  \textit{walk}(\textit{pos}
      {-}\textit{lbt}(l)){\cdot}\textit{Mule}(\textit{lbt}(l)) +{}$\\
&$\hspace{1.15cm}(\textit{lbt}(l){<}\textit{pos} {\wedge} \textit{pos} {<} \textit{rbt}(l)) \rightarrow{}$\\
&$\hspace{2cm}(\textit{walk}(\textit{pos}{+}\textit{rbt}(l){-}2{*}\textit{lbt}(l)) 
{\cdot}\textit{Mule}(\textit{rbt}(l)) +{}$\\
&$\hspace{2.15cm}\textit{walk}(2{*}\textit{rbt}(l){-}\textit{pos}{-} 
\textit{lbt}(l)){\cdot}\textit{Mule}(\textit{lbt}(l))));$\\

&\\
\textbf{init}& $\textit{Mule}(\textit{initial\_position});$\\
&\\
\hline
\end{tabular}
\end{center}
\caption{Woolhouse's model in mCRL2}
\label{figure:WoolhousemCRL2}
\end{figure}

Woolhouse's model in mCRL2 is given in Figure \ref{figure:WoolhousemCRL2}. 
The process $\textit{Mule}(\textit{pos})$ describes how the piecer repeatedly walks
back and forth the spinning mule to repair broken strands of yarn, where $\textit{pos}$ 
indicates the position where the piecer stands between two repair sessions. The action 
$\textit{threads}(l)$ with $l$ a list of threads, 
indicates the current threads of which $N$ are broken according to a uniform distribution. 
The position of the leftmost broken thread is indicated by $\textit{lbt}(l)$ and the rightmost broken thread by $\textit{rbt}(l)$.
Our piecer walks left, right-left, left-right, or only to the right to repair the broken strands.
The process allows the piecer to walk left-right or right-left without enforcing that the shortest distance is taken.
Using a quantitative modal formula, given below, we can investigate the shortest distance to be traversed. 
The action $\textit{walk}(d)$ indicates the distance $d$ that the piecer traverses, and the argument 
$\textit{pos}$ of $\textit{Mule}(\textit{pos})$ indicates
the position where the piecer ends.
\begin{figure}
\begin{center}
\begin{tabular}{|ll|}
\hline
&\\
\textbf{map}&$ \textit{Max}: \Pos;$\\
\textbf{eqn}&$\textit{Max}=50;$\\
&\\
\textbf{form}&$ 1/(\textit{Max}{*}\textit{width}){*}\nu X(n{:}\Nat=0).$\\
&$\hspace{3.5cm}            (n\dataeq \textit{Max} \wedge 0) \vee{}$\\
&  $\hspace{3.5cm}          (n<\textit{Max} \wedge
              [\true] ~\textbf{inf}~ d{:}\Int.[\textit{walk}(d)](\textit{d}{+}X(n{+}1)));$\\
&\\
\hline
\end{tabular}
\end{center}
\caption{The minimal distance covered by the piecer for $\textit{Max}$ repair rounds}
\label{fig:modalformula}
\end{figure}

\begin{table}[b]
\begin{center}
\begin{tabular}{|c|p{1.33cm}|p{1.33cm}|p{1.33cm}|p{1.33cm}|p{1.33cm}|p{1.33cm}|}
\hline
$N$&\multicolumn{6}{@{}c@{}}{
\begin{tabular}{|p{1.33cm}|p{1.33cm}|p{1.33cm}|p{1.33cm}|p{1.33cm}|p{1.33cm}|}
\multicolumn{6}{|p{8cm}|}{\mbox{Relative starting point of the piecer, i.e., $\textit{pos}/\textit{width}$}}\\
\hline
$0.0$&$0.1$&$0.2$&$0.3$&$0.4$&$0.5$\\
\end{tabular}}
\\
\hline
1   &0.4500  &0.3700    &0.3100     &0.2700    &0.2500    &0.2500    \\
2   &0.6333  &0.5711    &0.5378     &0.5244    &0.5222    &0.5222    \\
3   &0.7250  &0.6850    &0.6883     &0.7150    &0.7417    &0.7417    \\
4   &0.7800  &0.7600    &0.7929     &0.8524    &0.9057    &0.9057    \\
5   &0.8167  &0.8167    &0.8722     &0.9532    &1.0262    &1.0262    \\
6   &0.8429  &0.8629    &0.9362     &1.0291    &1.1148    &1.1148    \\
7   &0.8625  &0.9025    &0.9892     &1.0875    &1.1808    &1.1808    \\
8   &0.8778  &0.9378    &1.0333     &1.1333    &1.2311    &1.2311    \\
9   &0.8900  &0.9700    &1.0700     &1.1700    &1.2700    &1.2700    \\
10  &0.9000  &1.0000    &1.1000     &1.2000    &1.3000    &1.3000    \\
\hline
\end{tabular}
\end{center}
\caption{Walking distances relative to the width of the mule (model in Fig.~\ref{figure:WoolhousemCRL2}, $\textit{width}=10$, $\textit{Max}=1$)}
\label{table:mCRL2woolhouse}
\end{table}

In Figure \ref{fig:modalformula} a modal formula is given that provides the minimal average distance, relative to the
width of the mule, that the piecer needs to cover to do one round of repairs when he performs $\textit{Max}$ rounds
of repairs in total. Basically, it recursively calculates the cumulative distances $d$ in $\textit{Max}$ 
actions $\textit{walk}(d)$ and divides that by $\textit{Max}{*}\textit{width}$.
Our use of a maximal fixed point is of no relevance. It can be replaced by a minimal fixed point operator. Note that
$\vee$ stands for maximum and $\wedge$ for minimum.

When setting $\textit{width}=10$, we can redo Woolhouse's analysis by setting $\textit{Max}=1$. 
The results are listed in Table \ref{table:mCRL2woolhouse}. We make two observations. Compared to Table \ref{table:walking_distance}
with $\textit{width}=10000$, the distances in Table \ref{table:mCRL2woolhouse} are more extreme. This is caused by
the more even distribution of the broken threads when $\textit{width}$ is larger. Furthermore, columns $0.4$ and 
$0.5$ are equal, as with 10 strands they are symmetrical as no strands exist between threads 4 and 5. We like to stress
that with $\textit{Max}=1$ the walking distances predicted by the 
mCRL2 model coincides exactly with those of our probabilistic model of the Section \ref{sec:probmod}.

It is interesting to see what happens if our piecer repeatedly repairs strands. For this we let the piecer, rather arbitrarily,
repair the strands during 50 rounds, i.e., $\textit{Max}=50$ in the formula in Figure \ref{fig:modalformula}. The results are
listed in Table \ref{table:mCRL2woolhouserepeated}. We observe that, as expected, the initial position is hardly of relevance anymore,
and the expected walking distance of our piecer is slightly above the minimal distances of Woolhouse, but well
below the maximal distances. 
Especially, when a large
number of strands break, the piecer will automatically walk back and forth to optimise the distance. 
\begin{table}[h]
\begin{center}
\begin{tabular}{|c|p{1.33cm}|p{1.33cm}|p{1.33cm}|p{1.33cm}|p{1.33cm}|p{1.33cm}|}
\hline
$N$&\multicolumn{6}{@{}c@{}}{
\begin{tabular}{|p{1.33cm}|p{1.33cm}|p{1.33cm}|p{1.33cm}|p{1.33cm}|p{1.33cm}|}
\multicolumn{6}{|p{8cm}|}{\mbox{Relative starting point of the piecer, i.e., $\textit{pos}/\textit{width}$}}\\
\hline
$0.0$&$0.1$&$0.2$&$0.3$&$0.4$&$0.5$\\
\end{tabular}}
\\
\hline
1    &0.3324  &0.3308    &0.3296     &0.3288     &0.3284     &0.3284\\
2    &0.5664  &0.5652    &0.5646     &0.5645     &0.5646     &0.5646\\
3    &0.7091  &0.7083    &0.7083     &0.7088     &0.7093     &0.7093\\
4    &0.7898  &0.7894    &0.7900     &0.7910     &0.7920     &0.7920\\
5    &0.8358  &0.8358    &0.8369     &0.8383     &0.8396     &0.8396\\
6    &0.8634  &0.8638   &0.8653     &0.8670      &0.8685    &0.8685\\
7    &0.8807  &0.8815   &0.8832     &0.8851      &0.8867    &0.8867\\
8    &0.8916  &0.8928   &0.8947     &0.8967      &0.8984    &0.8984\\
9    &0.8978  &0.8994   &0.9014     &0.9034      &0.9053    &0.9053\\
10   &0.9000  &0.9020   &0.9040     &0.9060      &0.9080    &0.9080\\
\hline
\end{tabular}
\end{center}
\caption{Walking distances relative to the width of the mule (model in Fig.~\ref{figure:WoolhousemCRL2}, $\textit{width}{=}10$, $\textit{Max}{=}50$)}
\label{table:mCRL2woolhouserepeated}
\end{table}

\section{A more natural model}
\label{section3}
\begin{figure}[t]
\begin{center}
\begin{tabular}{|ll|}
\hline
\textbf{map}&$\textitprob: \Real;$\\
&     $\textit{probability}: \textit{List}(\textit{Thread})\rightarrow\Real;$\\
&\\
\textbf{eqn}&$  \textitprob=1/10;$\\
&\\
\textbf{var}&     $l: \textit{List}(\textit{Thread});$\\
\textbf{eqn}
&     $\textit{probability}([])=1;$\\
&     $\textit{probability}(\textit{fine} \triangleright l)=(1-\textitprob)*\textit{probability}(l);$\\
&     $\textit{probability}(\textit{broken} \triangleright l)=\textitprob*\textit{probability}(l);$\\
&\\
\textbf{proc}&$ \textit{Mule}(\textit{pos}:\Nat)=$\\
&       $\hspace*{0.5cm}\textbf{dist} ~l{:} \textit{List}(\textit{Thread})[\textit{if}(\#l{\dataeq} \textit{width}, \textit{probability}(l),0)].$\\
&       $\hspace*{1.5cm}     \textit{threads}(l){\cdot}$\\
&$  \hspace*{1.5cm}          (\textit{no\_broken\_thread}(l) )$\\
&$  \hspace*{1.9cm}          \rightarrow \textit{walk}(0){\cdot}\textit{Mule}(\textit{pos})$\\
&$    \hspace*{1.95cm}        \diamond ~~\,((\textit{pos}{\leq}\textit{lbt}(l))  \rightarrow  \textit{walk}(\textit{rbt}(l){-}\textit{pos}){\cdot}\textit{Mule}(\textit{rbt}(l)) +{}$\\
&$    \hspace*{2.5cm}            (\textit{pos}{\geq}\textit{rbt}(l)) \rightarrow  \textit{walk}(\textit{pos}{-}\textit{lbt}(l)){\cdot}\textit{Mule}(\textit{lbt}(l)) +{}$\\
&$    \hspace*{2.5cm}            (\textit{lbt}(l){<}\textit{pos} {\wedge} \textit{pos} {<} \textit{rbt}(l))$\\
&$    \hspace*{2.9cm}           \rightarrow{}   (\textit{walk}(\textit{pos}{+}\textit{rbt}(l){-}2{*}\textit{lbt}(l)){\cdot}\textit{Mule}(rbt(l)) +{}$\\
&$    \hspace*{3.53cm}              \textit{walk}(2{*}\textit{rbt}(l){-}\textit{pos}{-}\textit{lbt}(l)){\cdot}\textit{Mule}(\textit{lbt}(l))));$\\
&\\
\textbf{init}&$ \textit{Mule}(\textit{initial\_position});$\\
&\\
\hline
\end{tabular}
\end{center}
\caption{A more natural model in mCRL2}
\label{figure:naturalCRL2}
\end{figure}
Woolhouse's model assumes that the number of threads that break at each stroke is fixed.
But it seems more natural that every thread has an equal probability to break.
This can make quite a difference as sometimes no threads break, and at other times all threads are broken, with a substantial
impact on how far the piecer must walk on average. In this section we study this more natural model and find that this further
reduces the estimation of the distance the piecer must walk.

\subsection{A simple natural model}
In this section we present a model where each thread has a 
probability $\textitprob$ to break during a stroke of the mule. This model is depicted in Figure \ref{figure:naturalCRL2}. 
Note that we have omitted those data types and action declarations that are shared with the mCRL2
model in Figure \ref{figure:WoolhousemCRL2}. 

The essence of the model lies in the function $\textit{probability}$, that for a list of threads
calculates the likelihood of the pattern represented by the list occurring. This probability is an easy multiplication of the probabilities
$\textitprob$ for a thread that is $\textit{broken}$ and $1-\textitprob$ for a thread that is $\textit{fine}$.
The model chooses repeatedly a list of threads $l$ with the indicated probability. The 
piecer then walks from the current position $\textit{pos}$ to the leftmost broken thread ($\textit{lbt}(l)$) 
and the rightmost broken thread ($\textit{rbt}(l)$) either in a straight manner, or in a back and forward manner, except
if no threads are broken, in which case the piecer remains where he is and does not move.

\begin{table}[h]
\begin{center}
\begin{tabular}{|c|p{1.33cm}|p{1.33cm}|p{1.33cm}|p{1.33cm}|p{1.33cm}|p{1.33cm}|}
\hline
$N$&\multicolumn{6}{@{}c@{}}{
\begin{tabular}{|p{1.33cm}|p{1.33cm}|p{1.33cm}|p{1.33cm}|p{1.33cm}|p{1.33cm}|}
\multicolumn{6}{|p{8cm}|}{\mbox{Relative starting point of the piecer, i.e., $\textit{pos}/\textit{width}$}}\\
\hline
$0.0$&$0.1$&$0.2$&$0.3$&$0.4$&$0.5$\\
\end{tabular}}
\\
\hline
1    &0.2919  &0.2915    &0.2906     &0.2902     &0.2901     &0.2901\\
2    &0.5095  &0.5084    &0.5080     &0.5080     &0.5082     &0.5082\\
3    &0.6610  &0.6602    &0.6602     &0.6607     &0.6612     &0.6612\\
4    &0.7601  &0.7597    &0.7601     &0.7610     &0.7619     &0.7619\\
5    &0.8212  &0.8212    &0.8221     &0.8234     &0.8246     &0.8246\\
6    &0.8576  &0.8580    &0.8593     &0.8608     &0.8622     &0.8622\\
7    &0.8789  &0.8797    &0.8813     &0.8831     &0.8846     &0.8846\\
8    &0.8912  &0.8924    &0.8942     &0.8962     &0.8978     &0.8978\\
9    &0.8978  &0.8994    &0.9014     &0.9033     &0.9052     &0.9052\\
10   &0.9000  &0.9020    &0.9040     &0.9060     &0.9080     &0.9080\\
\hline
\end{tabular}
\end{center}
\caption{Walking distance for the natural model in Fig.~\ref{figure:naturalCRL2} ($\textit{width}=10$ and $\textit{Max}=50$)}
\label{table:simple_model}
\end{table}

In Table \ref{table:simple_model} we show the relative walking distance for our piecer as generated by the more
natural model for 50 strokes of a spinning mule of width 10. It should be noted that especially for smaller probabilities of a thread breaking,
the estimation is up to 10\% below that of Woolhouse. The computation of these probabilities
is expensive, as there are $2^{\textit{width}}$ breakage patterns to be considered for each stroke. In the next
section we show how this can be optimised. 

\subsection{An optimised natural model}
\begin{figure}[tp]
\begin{center}
\begin{tabular}{|ll|}
\hline
\textbf{map}&  $\textit{probability}: \Nat\times\Nat\rightarrow\Real;$\\
&\\
\textbf{var}&  $l,r: \Nat;$\\
\textbf{eqn}&  $\textit{probability}(l,r)=$\\
&$\hspace*{1cm}\textit{if}(l{<}\textit{width} \wedge r {<} \textit{width},$\\
&$\hspace*{2cm}    \textit{if}(l{<}r, \textit{exp}(1-\textitprob,l-1+\textit{width}-r)*\textitprob*\textitprob,$\\
&$\hspace*{2cm}                           \textit{if}(l{\dataeq}r,\textit{exp}(1{-}\textitprob,\textit{width}{-}1)*\textitprob,$\\
&$\hspace*{2cm}                           \textit{if}(l{+}1{\dataeq}\textit{width} {\wedge} r{\dataeq}0,\textit{exp}(1{-}\textitprob,\textit{width}),0))),0);$\\
&\\
\textbf{proc} &$\textit{Mule}(\textit{pos}:\Nat)=$\\
&\hspace*{0.5cm}$\textbf{dist }l,r{:} \Nat[\textit{probability}(l,r)].$\\
&\hspace*{1cm}$        \textit{threads}(l,r){\cdot}$\\
&\hspace*{1.5cm}$          ((l{>}r) \rightarrow \textit{walk}(0){\cdot}\textit{Mule}(\textit{pos}) +{}$\\
&\hspace*{1.65cm}$           (l{\leq}r {\wedge} \textit{pos}{\leq} l) \rightarrow \textit{walk}(r{-}\textit{pos}){\cdot}\textit{Mule}(r) +{}$\\
&\hspace*{1.65cm}$           (l{\leq}r {\wedge} l{<}\textit{pos} {\wedge} \textit{pos}{<}r) \rightarrow $\\
&\hspace*{2.1cm}$(\textit{walk}(\textit{pos}{+}r{-}2{*}l){\cdot}\textit{Mule}(r) + {}$\\
&\hspace*{2.25cm}$\textit{walk}(2{*}r{-}\textit{pos}{-}l){\cdot}\textit{Mule}(l)) +{}$\\
&\hspace*{1.65cm}$           (l{\leq}r {\wedge} \textit{pos}\geq r) \rightarrow \textit{walk}(\textit{pos}{-}l){\cdot}\textit{Mule}(l));$\\
&\\
\textbf{init }&$\textit{Mule}(\textit{initial\_position});$\\
\hline
\end{tabular}
\end{center}
\caption{An optimised natural model in mCRL2}
\label{figure:naturalmCRL2optimized}
\end{figure}

We optimise the model from the previous section using the observation that, for the purpose of analysing the distance traversed by the piecer, only the leftmost and rightmost
threads that are broken are relevant for our piecer. So, instead of exploring all possible patterns of broken threads we only 
want to know what the probability $\textit{probability}(l,r)$
is that the leftmost thread is at position $l$ and the rightmost thread is at position $r$.
The situation with no broken threads is represented by $r=0$ and $l=\textit{width}-1$, whereas the situation with exactly one broken thread is represented by $l = r$. The function $\textit{probability}(l,r)$
is given by:
\[\textit{probability}(l,r)=\left\{
\begin{array}{ll}
 (1{-}\textitprob)^\textit{width}           &\textrm{if }r=0~ \textrm{and}~ l=\textit{width}-1, \\ & \phantom{\textrm{if }}\textrm{i.e., no thread broken,}\\
 (1{-}\textitprob)^{(\textit{width}-1)}\textitprob     &\textrm{if }l=r, \\ & \phantom{\textrm{if }}\textrm{i.e., exactly one thread broken, and}\\
 (1{-}\textitprob)^{(l{-}1{+}\textit{width}-r)}\textitprob^2&\textrm{if }l<r,\\ & \phantom{\textrm{if }}\textrm{i.e., two or more threads broken.}\\
\end{array}\right.
\]
For the case two or more threads are broken, we observe that the probability has a factor $\textitprob^2$ because both the thread at position $l$ and the thread at position $r$ must be broken.  Furthermore, all threads before $l$ and after $r$ must be fine. There are $\textit{width}-r+l-1$ such
threads. All threads between positions $l$ and $r$ can be either broken or fine and therefore do not occur in the term.
Note that the probabilities for one and zero broken threads are independent of the values for $l$ and $r$. 

This leads to a straightforwardly adapted model which can be found in Figure~\ref{figure:naturalmCRL2optimized}. Again we leave
out those parts of the mCRL2 specification that can be found in the other places. 
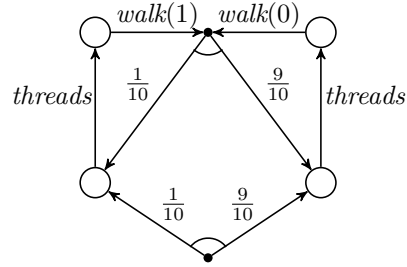
\begin{figure}[tp]
\begin{center}
\begin{tikzpicture}[> = stealth', semithick, inner sep = 1pt, scale=1]
  \tikzstyle{state} = [circle, draw=black, fill=white, minimum size=4.0mm]
  \tikzstyle{split} = [circle, draw=black, fill=black]
  \node [split] (t0) at (0,0) {};
  \node [state] (s0) at (-1.5,1) {$~~$};
  \node [state] (s1) at (1.5,1) {$~~$};
  \node [state] (s2) at (-1.5,3) {$~~$};
  \node [state] (s3) at (1.5,3) {$~~$};
  \node [split] (t1) at (0,3) {};
  \path [->] (t0) edge node [midway, above right] {$\frac{1}{10}$} (s0);
  \path [->] (t0) edge node [midway, above left] {$\frac{9}{10}$} (s1);
  \path [->] (s0) edge node [midway, above left] {$\textit{threads}$} (s2);
  \path [->] (s1) edge node [midway, above right] {$\textit{threads}$} (s3);
  \path [->] (s2) edge node [midway, above] {$\textit{walk}(1)$} (t1);
  \path [->] (s3) edge node [midway, above] {$\textit{walk}(0)$} (t1);
  \path [->] (t1) edge node [midway, above left] {$\frac{1}{10}$} (s0);
  \path [->] (t1) edge node [midway, above right] {$\frac{9}{10}$} (s1);

  \draw [>=] (t0) ++(31:.28) arc (217.5:322.5:-0.3);
  \draw [>=] (t1) ++(230:.3) arc (230:310:.3);

\end{tikzpicture}
\end{center}
\caption{The reduced probabilistic piecer's behaviour for the natural model ($\textit{width}=2$)}
\label{fig:bisimulation}
\end{figure}

Although we carefully derived the optimised model, and the calculations are not particularly difficult, we may want
to get extra assurance that the natural model and its optimisation are actually the same. Note that the parameters
of the action $\textit{threads}$ in both models are different and therefore we remove the parameters altogether so that 
the models are comparable.
Subsequently, we establish that the probabilistic state spaces of both models are strongly probabilistic bisimilar using the 
algorithm from \cite{GRV18}. This shows that the behaviour of the spinning mules, where we used width 10, is indeed 
equal. In order to get an impression of what such a state space looks like we depict the behaviour of a spinning mule of width 2
and a probability of $\frac{1}{10}$ that a thread breaks modulo strong probabilistic bisimulation in 
Figure~\ref{fig:bisimulation}. The probabilistic state at the bottom is
the initial state, which is bisimilar to the probabilistic state at the top. With probability $\frac{1}{10}$ the 
thread where the piecer is not standing breaks. This explains why he has to walk a distance 1 with probability $\frac{1}{10}$
whereas he can stand still, i.e., do the action $\textit{walk}(0)$, with probability $\frac{9}{10}$. As stated above 
the parameter of action $\textit{threads}$ has been removed.

\begin{figure}[tp]
\begin{center}
\begin{tikzpicture}
\draw [-, thick] (10.2,0) -- (-0.2,0) node[left] {$0.0$};
\draw [-, thick] (0,1) -- (-0.2,1) node[left] {$0.2$};
\draw [-, thick] (0,2) -- (-0.2,2) node[left] {$0.4$};
\draw [-, thick] (0,3) -- (-0.2,3) node[left] {$0.6$};
\draw [-, thick] (0,4) -- (-0.2,4) node[left] {$0.8$};
\draw [-, thick] (0,5) -- (-0.2,5) node[left] {$1.0$};
\draw [-, thick] (0,5.2) -- (0,0);
\draw [-, thick] (0,0) -- (0,-0.2) node[below] {$0.0$};
\draw [-, thick] (1,0) -- (1,-0.2) node[below] {$0.1$};
\draw [-, thick] (2,0) -- (2,-0.2) node[below] {$0.2$};
\draw [-, thick] (3,0) -- (3,-0.2) node[below] {$0.3$};
\draw [-, thick] (4,0) -- (4,-0.2) node[below] {$0.4$};
\draw [-, thick] (5,0) -- (5,-0.2) node[below] {$0.5$};
\draw [-, thick] (6,0) -- (6,-0.2) node[below] {$0.6$};
\draw [-, thick] (7,0) -- (7,-0.2) node[below] {$0.7$};
\draw [-, thick] (8,0) -- (8,-0.2) node[below] {$0.8$};
\draw [-, thick] (9,0) -- (9,-0.2) node[below] {$0.9$};
\draw [-, thick] (10,0) -- (10,-0.2) node[below] {$1.0$};
\draw [-, dashed, thick, red] 
(0,5*0.0000) -- 
(0.1,5*0.1568) -- 
(0.2,5*0.2871) -- 
(0.3,5*0.3968) -- 
(0.4,5*0.4883) -- 
(0.5,5*0.5644) -- 
(0.6,5*0.6272) -- 
(0.7,5*0.6789) -- 
(0.8,5*0.7214) -- 
(0.9,5*0.7562) -- 
(1,5*0.7848) -- 
(1.2,5*0.8280) -- 
(1.6,5*0.8794) -- 
(2,5*0.9074) -- 
(3,5*0.9414) -- 
(4,5*0.9573) -- 
(5,5*0.9665) -- 
(6,5*0.9723) -- 
(8,5*0.9783) -- 
(9,5*0.9795) -- 
(10,5*0.9800);   
\end{tikzpicture}
\end{center}
\caption{Relative walking distance $\textit{width}=50$, $\textit{Max}=50$ and $\textit{init}=0$}
\label{fig:graph}
\end{figure}
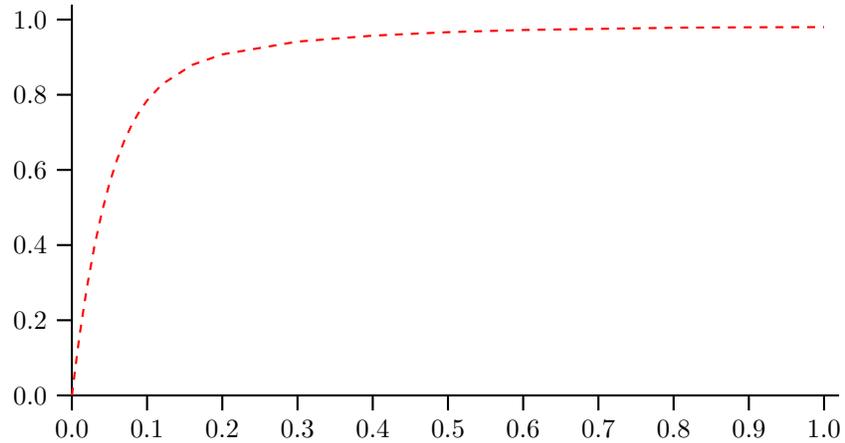

It is now possible to analyse this model for substantially larger widths of spinning mules. In Figure~\ref{fig:graph}
we depict the relative walking distance with increasing failure probability per thread for a mule of 50 threads wide. 
As the initial position is hardly of relevance, we only provide the numbers for the piecer starting at the leftmost position. As in Table~\ref{table:simple_model} we iterate 50 times.

We observe that, compared to Table~\ref{table:simple_model}, when on average the same number of threads break, this model of a wider mule predicts slightly shorter
walking distances for lower probabilities of breaking. 
That is, for one strand breaking on average per stroke, we compare the walking distance with 
$\textitprob=\frac{1}{10}$ for a mule of 10 threads wide, with that for $\textitprob=\frac{1}{50}$
for the mule that is 50 threads wide. The values are respectively, ${\geq}0.2901$ for a mule of 10 threads, whereas
this value is ${\leq}0.2871$ for a mule of 50 threads depending on the initial position of the piecer. 
This can be understood by noting that for mules with more threads, broken threads can be closer together.

The walking distances for higher probabilities of yarn to break increase with more strands. This is due to the
fact that the relative maximum walking distance for $\textit{width}$ threads is $\frac{\textit{width}-1}{\textit{width}}$,
which for $\textit{width}=10$ is $0.90$ and for $\textit{width}=50$ is $0.98$. If we were to compensate for this
the walking distance for lower breaking probabilities would reduce further. 

\section{Conclusion}
We modelled Woolhouse's Cotton-Spinning problem using a process formalism and analysed it using quantitative 
modal formulas. Our models are quite straightforward and describe the problem in a natural and
adequate way. 
Woolhouse concludes \emph{``the distances travelled (by piecers) exceeded thirty miles ($>$45km) per day''} \cite{Woolhouse64}.
Our models predict shorter walking distances for the piecers, and we must conclude that
Woolhouse's approach leads to an overestimation.

This seems in line with the observation by John Fielden, a British industrialist and Radical Member of Parliament for Oldham, around 1825 who stated in a speech
\emph{``At a meeting in Manchester a man claimed that a child in one mill walked twenty-four miles a day. I was surprised by this statement, therefore, when I went home, I went into my own factory, and with a clock before me, I watched a child at work, and having watched her for some time, I then calculated the distance she had to go in a day, and to my surprise, I found it nothing short of twenty miles''}.\footnote{\url{https://spartacus-educational.com/IRpiecers.htm}} 

Based on the data mentioned in the introduction, which stems from the early 20$^\textrm{th}$ century,
we assume 6 ruptured strands per 1,320 lines per 15 seconds, which
corresponds to $\textitprob=\frac{1}{220}$. Our last model gives a relative walking distance of $0.0761$,
which, for a mule of 46 meters wide, would lead to a walking distance of $8.4$km during a 10~hours working day. 
Such estimations depend on the quality of the machines and the cotton, as well as the width of the mule.
Although one must be careful to draw conclusions with such uncertainty of the elementary data, it appears
that working conditions, especially the average walking distances, of piecers 
in the early 20\textsuperscript{th} had improved based on technological improvement alone. 

It is interesting to reflect a little on the process description languages, quantitative modal logic 
and the supporting mCRL2 toolset. All models and the modal formula in this paper are available in the
example directory of the mCRL2 toolset (\url{www.mcrl2.org}). Most calculations have been carried out
with exact arithmetic, i.e., using precise fractions, which is very time consuming. The modal
formula and the respective process descriptions have been transformed to parameterised real equation 
systems (PRES) that have been solved using numerical approximation. This approach is very comparable to 
solving modal formulas via parameterised boolean equations (PBES, \cite{GWPBES,GM14}). As it stands the algorithms to 
solve PRESs are not very mature and are expected to substantially improve in the coming years. A particularly interesting 
avenue is to solve the PRESs symbolically, which is very successful for PBESs \cite{DBLP:conf/tacas/LaveauxWW22}, and which is also available in probabilistic model checking tools such as Storm~\cite{Storm} and Prism~\cite{Prism}.
Also a similar use of counter examples for PBESs in the context of PRESs may help to understand
how the models lead to particular numerical outcomes \cite{DBLP:conf/cade/WesselinkW18}.

\bibliographystyle{plain} 
\bibliography{references}

\end{document}